\documentclass[10pt,a4paper,onecolumn]{article}

  \usepackage{a4wide}

  \usepackage{hyperref}
  \usepackage{authblk}
 
  \providecommand{\tightlist}{%
  \setlength{\itemsep}{0pt}\setlength{\parskip}{0pt}}

  \NewDocumentCommand\citeproctext{}{}
  \NewDocumentCommand\citeproc{mm}{%
    \begingroup\def\citeproctext{#2}\cite{#1}\endgroup}
  \makeatletter
   \let\@cite@ofmt\@firstofone
   \def\@biblabel#1{}
   \def\@cite#1#2{{#1\if@tempswa , #2\fi}}
  \makeatother
  \newlength{\cslhangindent}
  \newlength{\csllabelwidth}
  \newenvironment{CSLReferences}[2] 
   {\begin{list}{}{%
    \setlength{\itemindent}{0pt}
    \setlength{\leftmargin}{0pt}
    \setlength{\parsep}{0pt}
    \ifodd #1
     \setlength{\leftmargin}{\cslhangindent}
     \setlength{\itemindent}{-1\cslhangindent}
    \fi
    \setlength{\itemsep}{#2\baselineskip}}}
   {\end{list}}
  \usepackage{calc}

  \title{SPM 25: open source neuroimaging analysis software}

    \author[1]{Tim M. Tierney}
    \author[1]{Nicholas A. Alexander}
    \author[1]{Nicole Labra Avila}
    \author[1]{Yaël Balbastre}
    \author[1]{Gareth Barnes}
    \author[1]{Yulia Bezsudnova}
    \author[1]{Mikael Brudfors}
    \author[1]{Korbinian Eckstein}
    \author[1]{Guillaume Flandin}
    \author[1]{Karl Friston}
    \author[1]{Amirhossein Jafarian}
    \author[1,4]{Olivia S. Kowalczyk}
    \author[1]{Vladimir Litvak}
    \author[1]{Johan Medrano}
    \author[1,5,6]{Stephanie Mellor}
    \author[1]{George O’Neill}
    \author[1,3]{Thomas Parr}
    \author[1,2]{Adeel Razi}
    \author[1]{Ryan Timms}
    \author[1]{Peter Zeidman}
    
    \affil[1]{Department of Imaging Neuroscience, University College London, UK}
    \affil[2]{School of Psychological Sciences and Turner Institute for Brain and Mental Health, Monash University, Clayton 3180, Australia}
    \affil[3]{Nuffield Department of Clinical Neurosciences, University of Oxford.}
    \affil[4]{Department of Neuroimaging, Institute of Psychiatry, Psychology \& Neuroscience, King’s College London}
    \affil[5]{Spinal Cord Injury Center, Balgrist University Hospital, University of Zurich, Zurich, Switzerland}
    \affil[6]{Translational Neuromodeling Unit, Institute for Biomedical Engineering, University of Zurich \& ETH Zurich, Zurich, Switzerland}

\begin{document}
  \maketitle

  \section{Summary}\label{summary}
  
  Statistical Parametric Mapping (SPM) is an integrated set of methods for
  testing hypotheses about the brain’s structure and function, using data
  from imaging devices. These methods are implemented in an open source
  software package, \texttt{SPM}, which has been in continuous development
  for more than 30 years by an international community of developers. This
  paper reports the release of \texttt{SPM\ 25.01}, a major new version of
  the software that incorporates novel analysis methods, optimisations of
  existing methods, as well as improved practices for open science and
  software development.
  
  \section{Statement of need}\label{statement-of-need}
  
  \texttt{SPM} introduced many of the statistical foundations that
  underpin cognitive and clinical neuroimaging research today, including:
  
  \begin{itemize}
  \tightlist
  \item
    The voxel-wise application of General Linear Models (GLMs) to
    neuroimaging data \\ (\citeproc{ref-friston1994statistical}{K. J.
    Friston, Holmes, et al., 1994}).
  \item
    Convolution modelling of functional MRI (fMRI) signals using
    haemodynamic response functions (\citeproc{ref-friston1994analysis}{K.
    J. Friston, Jezzard, et al., 1994}).
  \item
    Correction for multiple comparisons using topological inference
    (Random Field Theory, RFT) (\citeproc{ref-worsley1996unified}{Worsley
    et al., 1996}).
  \item
    Event-related fMRI (\citeproc{ref-josephs1997event}{Josephs et al.,
    1997}).
  \item
    Voxel-Based morphometry (VBM) for detecting changes in anatomy
    (\citeproc{ref-ashburner2000voxel}{Ashburner \& Friston, 2000}).
  \item
    Dynamic Causal Modelling (DCM) for state-space modelling using
    variational Bayesian methods (\citeproc{ref-friston2003dynamic}{K. J.
    Friston et al., 2003}).
  \item
    Source localisation for M/EEG data using variational Bayesian methods
    (\citeproc{ref-phillips2005empirical}{Phillips et al., 2005}).
  \end{itemize}
  
  These methods share certain key principles: the use of generative
  models, the application of well-motivated parametric statistics and a
  commitment to open science practices. They are included in a major new
  release of \texttt{SPM}, which addresses a series of needs in the
  neuroimaging community, set out below.
  
  \subsection{Open development}\label{open-development}
  
  \texttt{SPM} was previously developed and tested using a private
  Subversion server within University College London. To enable community
  engagement in the future development of \texttt{SPM} and to increase
  transparency, development has recently moved to a public
  \href{https://github.com/spm/spm}{GitHub repository}.
  \texttt{SPM\ 25.01} is the first release of the software following the
  move to GitHub. The key advantages of using GitHub thus far have been:
  
  \begin{itemize}
  \tightlist
  \item
    Introducing automated unit and regression tests across platforms.
  \item
    Automating the build process to conveniently generate and release
    source code and compiled versions.
  \item
    Issue tracking and distributing tasks among developers.
  \end{itemize}
  
  \subsection{Documentation and
  training}\label{documentation-and-training}
  
  The documentation for \texttt{SPM} was previously spread across multiple
  locations, most of which could not be edited by the community.
  \texttt{SPM\ 25.01} is accompanied by a new
  \href{https://www.fil.ion.ucl.ac.uk/spm/docs/}{documentation website},
  the source code for which is hosted in a public
  \href{https://github.com/spm/spm-docs}{GitHub repository}. The new
  website has step-by-step tutorials on all of SPM’s main features, as
  well as freely available video recordings of lectures from previous SPM
  courses covering the mathematical theory.
  
  \subsection{Major new features}\label{major-new-features}
  
  \texttt{SPM\ 25} includes 10 years of new developments since the last
  major release (\texttt{SPM\ 12}, dated 1st October 2014). This section
  highlights some of the most significant new features for different
  neuroimaging modalities.
  
  \subsubsection{MRI}\label{mri}
  
  \begin{itemize}
  \item
    Multi-Brain Toolbox (\citeproc{ref-brudfors2020flexible}{Brudfors et
    al., 2020}). Generates population average-shaped brains, enabling more
    precise spatial normalisation with the option to automatically label
    brain structures (\citeproc{ref-yan2022factorisation}{Yan et al.,
    2022}).
  \item
    SCOPE Toolbox. Generates voxel displacement maps (VDMs) using
    phase-encode-reversed pairs of MRI images (blip-up and blip-down
    images) to correct geometrical distortion in MRI
    (\citeproc{ref-andersson2003correct}{Andersson et al., 2003}). This is
    similar to the Topup toolbox in FSL.
  \end{itemize}
  
  \subsubsection{M/EEG}\label{meeg}
  
  \begin{itemize}
  \item
    Methods for spectral decomposition - \texttt{SPM\ 25.01} offers an
    implementation of an existing approach called FOOOF (Specparam) in the
    MEEGtools toolbox, based on code from Brainstorm
    (\citeproc{ref-donoghue2020parameterizing}{Donoghue et al., 2020}), as
    well as a new Bayesian implementation that introduces formal
    statistical testing, called Bayesian Spectral Decomposition (BSD)
    (\citeproc{ref-medrano2024bsd}{Medrano et al., 2024}).
  \item
    Support for fusion of different MEG sensor types and EEG sensors in
    beamforming with pre-whitening
    (\citeproc{ref-westner2022unified}{Westner et al., 2022}).
  \item
    Support for MEG BIDS for specification of events, channels and
    fiducials (\citeproc{ref-westner2022unified}{Westner et al., 2022}).
  \item
    Proof-of-concept routines for fusing M/EEG and fMRI data under a
    unified physiological model, to investigate neurovascular coupling
    (\citeproc{ref-friston2019dynamic}{K. J. Friston et al., 2019};
    \citeproc{ref-jafarian2019neurovascular}{Jafarian et al., 2019}).
  \end{itemize}
  
  \subsubsection{OPMs}\label{opms}
  
  A major recent innovation in neuroimaging is MEG using Optically Pumped
  Magnetometers (OPMs), which enable free movement of the head and body
  during neural recordings (\citeproc{ref-boto2018moving}{Boto et al.,
  2018}). This makes MEG available to new experimental paradigms (e.g.,
  experiments involving free movement
  (\citeproc{ref-mellor2023real}{Mellor et al., 2023})), new study
  populations who may not be amenable to traditional MEG (e.g., people
  with epilepsy (\citeproc{ref-mellor2024detection}{Mellor et al., 2024}))
  and recording of other biomagnetic fields (e.g., from the spinal cord
  (\citeproc{ref-spedden2024towards}{Spedden et al., 2024})). Developing
  analysis tools for OPM data is a major focus for SPM, with recently
  added features including:
  
  \begin{itemize}
  \tightlist
  \item
    File IO for all major OPM manufacturers (Quspin, Cerca, Mag4Health,
    Fieldline).
  \item
    Methods to simulate arbitrary OPM arrays of differing densities and
    vector measurements.
  \item
    OPM interference cancellation algorithms for low channel systems:
    Homogeneous Field Correction
    (\citeproc{ref-tierney2021modelling}{Tierney et al., 2021}).
  \item
    OPM interference cancellation algorithms for large channel systems:
    Adaptive Multipole Models (\citeproc{ref-tierney2024adaptive}{Tierney
    et al., 2024}).
  \end{itemize}
  
  \subsubsection{Bayesian statistics}\label{bayesian-statistics}
  
  \begin{itemize}
  \item
    Parametric Empirical Bayes (PEB)
    (\citeproc{ref-friston2016bayesian}{K. J. Friston et al., 2016})
    extends the Dynamic Causal Modelling (DCM) framework to include random
    effects modelling of neural connectivity parameters, enabling people
    to test hypotheses about the similarities and differences among
    research participants.
  \item
    Bayesian model reduction (BMR) (\citeproc{ref-friston2018bayesian}{K.
    Friston et al., 2018}) enables statistical evidence to be rapidly
    scored for large numbers of competing models, where models differ only
    in their priors.
  \end{itemize}
  
  \subsubsection{Behavioural modelling}\label{behavioural-modelling}
  
  In addition to neuroimaging analysis, SPM includes a suite of tools for
  behavioural modelling, including a comprehensive repository for
  computational neuroscience using the Active Inference framework. The
  code in \texttt{SPM\ 25.01} has undergone significant development,
  offering a range of demonstrations accessible via the SPM DEM toolbox
  and associated GUI, and detailed in an accompanying textbook
  (\citeproc{ref-parr2022active}{Parr et al., 2022}). The key features
  are:
  
  \begin{itemize}
  \tightlist
  \item
    A series of inversion schemes for generative models based upon
    Partially Observable Markov Decision Processes (POMDPs) that can be
    used to simulate sequential choices, decision making, and planning
    (\citeproc{ref-friston2017active}{K. Friston et al., 2017}).
  \item
    An active (generalised) filtering scheme for numerical simulation of
    continuous movement behaviour and responses to continuous sensory
    signals, e.g., (\citeproc{ref-friston2015duet}{K. Friston \& Frith,
    2015}).
  \item
    Options for hierarchical composition of the above models
    (\citeproc{ref-friston2017graphical}{K. J. Friston et al., 2017}) and
    composition with a range of other models (e.g., speech recognition
    (\citeproc{ref-friston2021active}{K. J. Friston et al., 2021})).
  \item
    Routines to fit the above models to behavioural data
    (\citeproc{ref-schwartenbeck2016computational}{Schwartenbeck \&
    Friston, 2016}).
  \end{itemize}
  
  \subsubsection{SPM without MATLAB}\label{spm-without-matlab}
  
  Approximately 90\% of the \texttt{SPM\ 25.01} source code is written in
  MATLAB and the remainder is written in C. This code has been highly
  optimised and thoroughly tested over 30 years of development. We have
  therefore carefully considered how to capitalise on the stability of the
  \texttt{SPM} software, while making it more accessible for people who do
  not have access to a MATLAB license, or who prefer to write their
  analysis code in other languages.
  
  Our strategy is as follows:
  
  \begin{itemize}
  \item
    \texttt{SPM\ 25} will be the first version of SPM to be fully
    accessible from the Python programming language, without requiring
    MATLAB, using a new Python wrapper called
    \href{https://github.com/spm/spm-python}{spm-python}. This is in the
    final stages of development and will be released in the first quarter
    of 2025.
  \item
    \href{https://www.fil.ion.ucl.ac.uk/spm/docs/installation/standalone/}{SPM
    Standalone} is the compiled version of \texttt{SPM} that can be run
    from the command line without a MATLAB license. This enables people to
    run neuroimaging analyses from command line scripts written in any
    language, or using the GUI. It is now generated automatically with
    each new release, as part of the GitHub-based build process.
  \item
    \href{https://www.fil.ion.ucl.ac.uk/spm/docs/installation/containers/}{Docker
    and Singularity containers} are additionally provided and are now
    generated automatically as part of SPM’s GitHub build process.
  \end{itemize}
  
  \section{Software versions}\label{software-versions}
  
  \texttt{SPM\ 25.01} is the first release of \texttt{SPM} to use calendar
  versioning, thus \texttt{SPM\ 25.01} is the version issued in January
  2025. All releases are available via
  \url{https://github.com/spm/spm/releases}.
  
  \section{Acknowledgements}\label{acknowledgements}
  
  A full list of authors of \texttt{SPM} can be found in the file
  \texttt{AUTHORS.txt} supplied with the software. We are also grateful to
  the IT Team at the UCL Department of Imaging Neuroscience for their
  ongoing support.
  
  Yael Balbastre is funded by a Royal Society Newton International
  Fellowship (NIF\textbackslash R1\textbackslash232460). Olivia S.
  Kowalczyk is supported by the King’s Prize Fellowship. Nicholas A
  Alexander and Johan Medrano are supported by the Discovery Research
  Platform for Naturalistic Neuroimaging funded by Wellcome
  {[}226793/Z/22/Z{]}. Stephanie Mellor was funded by an Engineering and
  Physical Sciences Research Council (EPSRC) Healthcare Impact Partnership
  Grant (EP/V047264/1). Tim M. Tierney is funded by an Epilepsy Research
  UK fellowship (FY2101). Thomas Parr is supported by NIHR Academic
  Clinical Fellowship (ref: ACF-2023-13-013). Peter Zeidman is funded by
  an MRC Career Development Award {[}MR/X020274/1{]}.
  
  \section*{References}\label{references}
  \addcontentsline{toc}{section}{References}
  
  \phantomsection\label{refs}

  \end{document}